\documentclass[12pt,fleqn]{article}
\usepackage{amsfonts,ulem,url}
\usepackage{amssymb}
\usepackage{amsmath}
\usepackage{amstext,verbatim,graphicx,ifthen,multirow,amsthm}
\usepackage{bm}
\usepackage{booktabs}

\usepackage{natbib}

\renewcommand{\mathbf}{\boldsymbol}
\renewcommand{\thepage}{}
\renewcommand{\appendix}{\footnotesize\parindent 0cm\setcounter{equation}{0}
\renewcommand{\theequation}{A.\arabic{equation}}
\setcounter{lemma}{0}\renewcommand{\thelemma}{A.\arabic{lemma}}}

\newcommand{\tick}{\checkmark}
\newcommand{\tock}{?}

\newcommand{\oy}{\overline{Y}}

\newcommand{\imp}{{\rm imp}}

\newcommand{\ppo}{{\rm O}}
\newcommand{\pp}{{\rm P}}
\newcommand{\pps}{{\rm S}}
\newcommand{\ppe}{{\rm E}}

\newcommand{\weight}{{\rm weight}}

\newcommand{\been}{\mathbf{1}}

\newcommand{\mmx}{\mathbb{X}}

\newcommand{\mme}{\mathbb{E}}

\newtheorem{assumption}{Assumption}
\newtheorem{theorem}{Theorem}
\newtheorem{lemma}{Lemma}

\newtheorem{remark}{Remark}
\newcommand{\indep}{\perp\!\!\!\perp}

\def\monthname{\ifcase\month\or
  January\or February\or March\or April\or May\or June\or July\or
  August\or September\or October\or November\or December\fi}
\renewcommand{\thepage}{[\arabic{page}]}
\numberwithin{equation}{section}

\def\monthname{\ifcase\month\or
January\or February\or March\or April\or May\or June\or
July\or August\or September\or October\or November\or December\fi}
\renewcommand{\appendix}{\small\parindent 0cm\setcounter{equation}{0}
\renewcommand{\theequation}{A.\arabic{equation}}
\setcounter{lemma}{0}\renewcommand{\thelemma}{A.\arabic{lemma}}
\setcounter{theorem}{0}\renewcommand{\thetheorem}{A.\arabic{theorem}}}
\begin{document}

\title{
Using Experiments to Correct for Selection in Observational Studies\thanks{{\small We are
grateful for discussions with Nathan Kallus, Xiaojie Mao, Dylan Small, and for comments from seminar participants at the University of Pennsylvania and Berkeley. We also want to acknowledge excellent research assistance from Kevin Chen. 
 This research was funded
through the Sloan Foundation, Schmidt Futures, and ONR grant N00014-17-1-2131.}} }
\author{Susan Athey\thanks{{\small Graduate School of Business, Stanford University, and NBER,
athey@stanford.edu. }} \and Raj Chetty\thanks{{\small Department  of
Economics, Harvard University, and NBER,
chetty@harvard.edu. }}
 \and Guido W. Imbens\thanks{{\small 
Graduate School of Business, and Department of Economics, Stanford University, and NBER,
imbens@stanford.edu. }} 
 }
\date{First Version, August 2019;  Current version \ifcase\month\or
January\or February\or March\or April\or May\or June\or
July\or August\or September\or October\or November\or December\fi \ \number%
\year\ \  }
\maketitle

\begin{abstract}
In the social sciences there has been an increase in interest in randomized experiments to estimate causal effects during the last twenty years, partly  because their internal validity  tends to be high, but they are often small and contain information on only a few variables. At the same time, as part of the big data revolution, large, detailed, and  representative, administrative data sets have become more widely available. However, the credibility of estimates of causal effects based on such data sets alone can be low.
 In this paper, we develop statistical methods for systematically combining experimental and observational data to improve the credibility of estimates of the causal effects. We focus on a setting with a binary treatment where we are interested in the effect on a primary outcome that we only observe in the observational sample. Both the observational and experimental samples contain data about a treatment, observable individual characteristics, and a secondary (often short term) outcome.  To estimate the effect of a treatment on the primary outcome, while accounting for the potential confounding in the observational sample, we propose a method that makes use of estimates of the relationship between the treatment and the secondary outcome from the experimental sample. We interpret differences in the estimated causal effects on the secondary outcome between the two samples as evidence of unobserved confounders in the  observational sample, and develop control function methods for using those differences to adjust the estimates of the treatment effects on the primary outcome. 
 We illustrate these ideas by combining data on class size and third grade test scores from the  Project STAR experiment  
 with observational data  on class size and both third and eighth grade test scores from the New York school system. 
\end{abstract}




\baselineskip=20pt\newpage \setcounter{page}{1}\renewcommand{\thepage}{[%
\arabic{page}]}\renewcommand{\theequation}{\arabic{section}.%
\arabic{equation}}


\section{Introduction}
\label{section:introduction}

There has been an influential movement in empirical studies in economics towards relying more on experimental as opposed to observational data to estimate causal effects ({\it e.g.}, \citet{duflo2007using, angrist2010credibility}). 
The internal validity of randomized experiments tends to be high, and their analyses are relatively straightforward (\citet{athey2017econometrics}). However, due to the practical challenges involved in running experiments ({\it e.g., }\citet{glennerster2013running}), there are often drawbacks to using experimental data. Experiments are often limited in sample size, in the richness of the information collected, as well as in representativeness,
raising concerns about external validity. At the same time, as part of the big data revolution, large, detailed, and by their nature often representative, administrative data sets have become more widely available ({\it e.g.}, \citet{chetty}). However, it is challenging to use such datasets to estimate causal effects because observational studies often lack internal validity. In this paper, we develop statistical methods for systematically combining experimental and observational data in an attempt to leverage the strengths of both types of data. We focus on a canonical case where both experimental and observational data contain information about individual treatment assignments and a secondary (e.g. short term) outcome (where the datasets contain different individuals), but only  the observational data  contains information about the primary (often long term) outcome of interest.

We illustrate our methods combining data from the New York school system (the ``observational sample'') and from Project STAR (the ``experimental sample'', see \citet{krueger2001effect} for an earlier analysis).  Our goal is to estimate the effect of class size on 8th grade test scores in New York (what we call the ``primary outcome''). However, these 8th grade test scores are not available in the Project STAR data; instead, the experimental sample includes test scores only through the 3rd grade (what we call the ``secondary outcome'').  See Table 1 for the average outcomes in each sample by treatment status.
\begin{table}[tbh]
        \caption{Average Outcomes and Standard Errors  by Sample and Treatment Status}
        \label{tab:}
        \centering
        \vspace{1em}
        \begin{tabular}{lccccc}
\toprule
{} &                                     Project STAR &                                \multicolumn{2}{c}{New York} \\
{} &                                      3rd Grade Score &                                3rd Grade Score  &                                         8th Grade Score \\
{} &                                      (secondary outcome) &                             (secondary outcome)  &                                           (primary outcome) \\
\midrule\\
Mean Controls (regular class size)           &                      $0.011$\textsuperscript{} &                     $0.157$\textsuperscript{} &                     $0.155$\textsuperscript{} \\
                 &  ($0.015$\textsuperscript{})\textsuperscript{} &  ($0.001$\textsuperscript{})\textsuperscript{} &  ($0.001$\textsuperscript{})\textsuperscript{} \\ \\
 Mean Treated    (small class size)        &                      $0.193$\textsuperscript{} &                     $0.070$\textsuperscript{} &                     $-0.028$\textsuperscript{} \\
                 &  ($0.025$\textsuperscript{})\textsuperscript{} &  ($0.001$\textsuperscript{})\textsuperscript{} &  ($0.002$\textsuperscript{})\textsuperscript{} \\ \\
  Difference            &                      $0.181$\textsuperscript{} &                     $-0.087$\textsuperscript{} &                     $-0.183$\textsuperscript{} \\
                 &  ($0.029$\textsuperscript{})\textsuperscript{} &  ($0.002$\textsuperscript{})\textsuperscript{} &  ($0.002$\textsuperscript{})\textsuperscript{} \\ \\
                 
\bottomrule
\end{tabular}
        \end{table}
        We find that, even after adjusting for observed pre-treatment student characteristics, the estimated effect of class size on third grade scores is quite different in Project STAR than it is in the observational data from New York.
         For the experimental sample from Project STAR we see that there is a substantial positive effect of the small class size, an increase of 0.181 in 3rd grade scores. On the other hand, in the observational sample from New York we see a substantial negative relationship between the treatment and test scores in both 3rd and 8th grade, -0.087 for 3rd grade scores and -0.183 for 8th grade scores. 
         
In  this paper we interpret the difference in 3rd grade score estimates, 0.181 and -0.087 as the result of unobserved confounders present in the observational sample. We use the distribution of outcomes in the experimental sample to disentangle the contribution of the unobserved confounders in the observational sample.         Under the strong assumption, latent unconfoundedness, that the unobserved confounders for the primary and secondary outcome are the same, we can use this to obtain credible estimates of the causal effects on the primary outcome.

We can interpret this difference in the 3rd grade tests results,  0.181 in Project STAR {versus} -0.087 in New York, in two ways. One interpretation is that the difference (even after adjusting for pre-treatment variables) is due to differences between the two populations, so that it reflects lack of external validity of the Project STAR sample. A second interpretation is that it reflects lack of internal validity or non-random selection into the treatment in New York, in other words, the presence of unobserved confounders in the New York sample. 
(In practice, it is of course possible that both complications are present.)
In this paper we focus on the latter explanation, and maintain the assumption that the experimental dataset has both internal and external validity; that is, we assume that after adjusting for pre-treatment variables, the underlying populations in the experimental and observational datasets are comparable, even if treatments are assigned differently. Under this maintained assumption, the 0.181 is the preferred estimate of the causal effect on small classes on 3rd grade scores.  Given that interpretation  the negative correlation between the treatment and 3rd grade outcomes in New York must be due to unobserved confounding, for example, sorting of students who are likely to test poorly into schools with low class sizes. As a result, it appears implausible to interpret the correlation between 8th grade scores and class size in New York as causal.

The main question we address in this paper is how we can adjust the 8th grade results for New York in Table 1 in the light of the experimental Project STAR 3rd grade results and the New York 3rd grade results, under the assumption that the difference in 3rd grade results is due to endogenous selection into the treatment or lack of internal validity in New York. 
Because 8th and 3rd scores may be measured on different scales, we cannot simply subtract the difference in 3rd grade effects in the two samples, in a difference-in-differences approach.
Instead our approaches uses 
the relation between 3rd grade scores and class size in New York (which is a combination of the causal effect and the selection effect) and the relation between the 3rd grade scores and class size in Project Star (which is causal) to extract the selection component. We then adjust for this selection component to remove the endogeneity in the relation between 8th grade scores and class size in New York.


Formally, we consider a set up with two datasets,  the experimental sample and  the observational sample. For each unit in the observational dataset, we observe pre-treatment variables, a binary treatment assignment, the primary outcome, and a (vector-valued) secondary outcome.  For each unit in the experimental dataset, we observe the same variables as in the observational dataset, except that we do not observe the primary outcome.  Table \ref{tabel1} illustrates this observational scheme. This observation scheme is also studied in
\citet{rosenman2018propensity, rosenman2020combining, kallus2020role}.
\citet{rosenman2018propensity} focuses on the problem where assignment is unconfounded in both samples.
\citet{kallus2020role} considers the case where assignment in the combined sample is unconfounded, but not in each of the samples separately.
\citet{rosenman2020combining} allow for unobserved confounders in the observational sample and consider shrinkage estimators.
\citet{kallus2018removing} focus on a different case where the same variables are observed in the two samples, but as in our set up, unconfoundedness does not hold in the observational sample. 
\citet{mealli2013using} focuses on an instrumental variables setting where the presence of multiple outcomes can improve estimates.

This set up  in this paper differs from the  surrogate set up in \citet{athey2019surrogate} where in the observational study the treatment indicator $W_i$ is not observed. The fact that in the observational sample the treatment  indicator is observed allows us to relax the surrogacy assumptions \citet{athey2019surrogate} exploit, in particular the assumption that the only effect of the treatment on the primary outcome is through the secondary, intermediate, outcome.
Typically, in our setting, the primary outcome is a long-term outcome such as eventual educational attainment, long-term wages, or mortality, while the secondary outcome may be a multi-dimensional vector of shorter-term outcomes that are associated with the long-term outcome.  The object of interest is a low-dimensional estimand, for example, the average causal effect of the treatment on the primary outcome. The role of the secondary outcome and the pretreatment variables is to aid in the effort of credibly estimating the average causal effect on the primary outcome.

\noindent 
\begin{table}\label{tabel1}
\begin{centering}
\textsc{Table 1. Observation Scheme: $\tick$ is observed, }\tock\textsc{ is missing}
\par\end{centering}
\centering{}\label{tabel_sampling}\vskip0.3cm \centering{}%
\begin{tabular}{c|ccccc}
 &  &  & Primary  &  Secondary & Pretreatment \\
 & Sample &Treatment  & Outcome & Outcome  &  Variables\tabularnewline
Units & $G_{i}$ & $W_{i}$ & $Y^\pp_{i}$ & $Y^\pps_{i}$ & $X_{i}$\tabularnewline
 &  &  &  &  & \tabularnewline
\hline 
 &  &  &  &  & \tabularnewline
1 to $N_{\ppe}$  & $\ppe$ & \tick & \tock & \tick & \tick\tabularnewline
 &  &  &  &  & \tabularnewline
$N_{\ppe}+1$ to $N_{\ppe}+N_{\ppo}$ & $\ppo$ & \tick & \tick & \tick & \tick\tabularnewline
\end{tabular}
\end{table}

Our approach makes use of three maintained assumptions.
First, the sample of units in the observational dataset is representative of the population of interest.
This assumption essentially  defines the estimand.  However, the problem is that treatment is not randomly assigned in the observational data. 
Second, the treatment in the experimental study was  randomly assigned, ensuring  that the experimental study has internal validity. We can easily generalize this to the case where the maintained assumption is that treatment assignment in the second sample is unconfounded given a set of pretreatment variables (\citet{rosenbaum1983central, imbens2015causal}). 
In our application this assumption is satisfied by design.
However, because the primary outcome is not observed in this data set, we cannot estimate the average effect of interest on the experimental sample alone. Third,  we assume that the pretreatment variables capture the differences between the populations that the observational and experimental sample were drawn from, so that conditional on these pretreatment variables, estimates of the treatment effect in the experimental sample have external validity (\citet{shadishcookcampbell, hotz2005predicting}). 
 However, these three maintained assumptions, external validity for the observational sample, and internal validity and conditional external validity for the experimental sample, are not sufficient for identification of the average causal effect of the treatment on the primary outcome.

Our first and main contribution to this problem is to formulate a novel assumption, which we call ``latent unconfoundedness.'' In combination with the three maintained assumptions, this  latent unconfoundedness assumption allows for 
point-identification of the average causal effect of the treatment on the primary outcome in the observational study, without generating testable implications. 
The latent unconfoundedness  assumption implies that the
the unobserved confounders  in the observational sample that confound treatment assignment and the secondary outcome are the same unobserved confounders that affect both treatment assignment and the primary outcome. Moreover, observing  the secondary outcome allows for the extraction of this confounder by exploiting the presence of the experimental data.
Formally the assumption links (without the need for functional form assumptions) the biases in treatment-control differences in the secondary outcome (which can be estimated given the presence of the experimental data) to the biases in treatment-control comparisons in the primary outcome (which the experimental data are silent about) using  a control function approach (\citet{heckman1979sample, heckman1985alternative, imbens2009identification, wooldridge2010econometric}). This approach also bears some similarity to the Changes-In-Changes approach in \citet{athey2006identification}. For a unit in the observational sample the control function is essentially the rank of the secondary outcome in the distribution of secondary outcomes in the experimental sample with the same treatment.
The method makes use of the fact that under our maintained assumptions, systematic differences in the estimated effect of the treatment between the experimental and observational sample are attributed to violations of unconfoundedness in the observational data.

 In our second contribution, we propose three different approaches to estimation of  the average treatment effect under the maintained assumptions in combination with latent uconfoundedness.
 The three approaches consist of $(i)$ imputating the missing primary outcome in the experimental sample, $(ii)$ weighting of the units in the observational sample to remove biases, and $(iii)$ control function methods.
Our analyses show how the presence of the experimental data can be systematically exploited to relax the assumption of unconfoundedness that is common in observational studies.

In our third contribution we apply the new methods to obtain estimates of the effect of small class sizes on 8th grade test scores in New York. The combination of the New York data with the experimental Project STAR data leads to a positive estimates of the effect of small classes. In contrast,  an analysis using only the New York data, and assuming unconfoundedness, leads to implausible negative estimates.

\section{Two  Examples}\label{two_examples}

To lay out the conceptual issues at the heart of the current paper we considerin this section  two simple examples  in some detail. These two examples allow us to introduce the identifying assumptions and estimation strategies that are the main contribution of this paper.

\subsection{Set Up}

The basic set up is the same in both examples.
 Using the potential outcome set up developed for observational studies by \citet{rubin1974estimating} (see \citet{imbens2015causal} for a textbook discussion), let the pair of potential outcomes for this outcome for unit $i$ 
be denoted by $Y_{i}^\pp(0)$
and $Y_{i}^\pp(1)$,
where the superscript ``$\pp$'' stands for ``Primary''. In many applications this is a long term outcome.
In our application this is 8th grade scores.
 The treatment received by unit $i$ will be denoted by  $W_{i}\in\{0,1\}$, an indicator for small class size in our application. There is also a secondary outcome, possibly a short term outcome, with the pair of potential outcomes for unit $i$ denoted  by $Y_{i}^\pps(0)$
  and $Y_{i}^\pps(1)$, where the superscript ``$\pps$'' stands for ``Secondary''. In our application this is a 3rd grade score.
  In the two examples both the primary and secondary outcomes are scalars, but in applications the secondary outcome is likely to be vector-valued.
The realized values for the primary and secondary outcomes are $Y^\pp_i\equiv Y_i^\pp(W_i)$ and  $Y^\pps_i\equiv Y_i^\pps(W_i)$.  We may also observe pretreatment variables, denoted by $X_i$

We are interested in the average treatment effect on the primary outcome,
\begin{equation}\label{taup} \tau^\pp\equiv\mme\left[ Y^\pp_i(1)-Y^\pp_i(0)\right],\end{equation}
although other estimands such as the average effect on the treated can be accomodated in this set up.
The average effect on the secondary outcome, $ \tau^\pps\equiv \mme\left[ Y^\pps_i(1)-Y^\pps_i(0)\right],$
is, for the purpose of the current study, not of intrinsic interest.

We have two samples to draw on for estimation of $\tau^\pp$. In that sense the set up connects to the literature on combining data sets, {\it e.g.,} \citet{hotz2005predicting, pearl2014external, ridder2007econometrics}. The first sample is from an observational study. It is a random sample from the population of interest.   For all units in this observational sample we observe the quadruple $(W_i,Y^\pps_i,Y^\pp_i,X_i)$, 
The second sample is a possibly  selective sample from the same population, with the assignment completely random. 
 For all units in  this experimental sample we observe the triple $(W_i,Y^\pps_i,X_i)$, but not the primary outcome. 
 The motivation for considering this setting is that it is often expensive to conduct randomized experiments, and it may not be feasible to observe the primary outcome in the experiment.

 Let 
  $G_i\in\{\ppe,\ppo\}$, be the indicator for the subpopulation or group a unit is drawn from. Then
   we can think of the combined sample as a random sample of size $N$ from an artificial super-population for which we observe the quadruple
$(W_i,G_i,Y^\pps_i,Y^\pp_i\been_{G_i=\ppo},X_i)$, where $\been_{G_i=\ppo}$ is a binary indicator, equal to 1 if $G_i=\ppo$ and equal to 0 if $G_i=\ppe$.

\subsection{A Binary Outcome Example}

For the purpose of the first example in this section, we assume both the secondary and primary outcome are binary, $Y^\pp_i(w),Y^\pps_i(w)\in\{0,1\}$ for $w\in\{0,1\}$.
For expositional reasons we also assume in this section that there are no pretreatment variables.

For all outcome types  $t\in\{\pps,\pp\}$, all groups  $g\in\{\ppe,\ppo\}$, and all treatment levels  $w\in\{0,1\}$ the sample averages and sample sizes, define
\[ \oy^{t,g}_w\equiv \frac{1}{N^g_w}\sum_{i=1}^N Y^t_i \been_{G_i=g,W_i=w},
\hskip1cm {\rm and}\ \ 
N^g_w\equiv  \sum_{i=1}^N  \been_{G_i=g,W_i=w}.\]
Assuming that  $N^\ppo_0$, $N^\ppo_1$,  $N^\ppe_0$, and $N^\ppe_1$ are all positive, 
 six of the eight average outcomes $\oy^{t,g}_w$ are well-defined and can be calculated from the data,
$\oy^{\pp,\ppo}_0$, $\oy^{\pp,\ppo}_1$, $\oy^{\pps,\ppo}_0$, $\oy^{\pps,\ppo}_1$, $\oy^{\pps,\ppe}_0$, and $\oy^{\pps,\ppe}_1$. The remaining two, 
 $\oy^{\pp,\ppe}_0$ and $\oy^{\pp,\ppe}_1$ are not  well-defined because we do not observe the primary outcome in the experimental sample.

\subsubsection{Using the Two Samples Separately}

Let us first consider estimation of $\tau^\pp$ and $\tau^\pps$ using one sample at a time.
Using only the experimental sample there is no way to estimate the average treatment effect on the primary outcome, because this sample does not contain any information on the primary outcome. 
We can estimate the average effect on the secondary outcome, using the experimental sample, as
\[ \hat\tau^{\pps,\ppe}=\oy^{\pps,\ppe}_1-\oy^{\pps,\ppe}_0.\]
This estimator $\hat\tau^{\pps,\ppe}$  would be unbiased for $\tau^\pps$ if the experimental sample had external validity and could be considered  a random sample from the population of interest (formally, if $G_i\indep (Y_i(0),Y_i(1))$, what \citet{hotz2005predicting} call location unconfoundedness). Without that assumption this estimator would not necessarily be unbiased.

Using only the observational sample the natural estimator for the average causal effect on the primary and secondary outcomes would be
\[ \hat\tau^{\pp,\ppo}=\oy^{\pp,\ppo}_1-\oy^{\pp,\ppo}_0,\hskip1cm {\rm and}\ \ 
 \hat\tau^{\pps,\ppo}=\oy^{\pps,\ppo}_1-\oy^{\pps,\ppo}_0
,\]
respectively.
For these estimators to be consistent for the average causal effect of the treatment on the primary and secondary outcomes we would need something like unconfoundedness, which, in the absence of pretreatment variables,  corresponds to:
\[ W_i\ \indep\ \Bigl(Y^\pps_i(0),Y^\pps_i(0),Y^\pp_i(0),Y^\pp_i(0)\Bigr)\ \Bigr|\ G_i=\ppo.\]
With only the observational sample, there is not much  in terms of alternatives for obtaining point estimates of the average treatment effect on the primary and secondary outcomes.

\subsubsection{Combining the Two Samples}

Now consider estimation of $\tau^\pps$ in the presence of both experimental and observational samples.  
In this case we have two distinct estimators for $\tau^\pps$, namely
$\hat\tau^{\pps,\ppe}$ and $\hat\tau^{\pps,\ppo}$. 
If we find no difference between $\hat\tau^{\pps,\ppe}$ and $\hat\tau^{\pps,\ppo}$, or at least no statistically significant difference, then both would appear to be reasonable estimates.
We might improve the precision of either estimator by combining them efficiently ({\it e.g.,} \citet{rosenman2018propensity, rosenman2020combining, kallus2020role}).
However, if we find a substantial and statistically significant difference between the two, 
  we can infer that either  the assignment in the observational sample is not  random (unconfoundedness does not hold), or the experimental sample is not a random sample from the population of interest (no external validity). 
  In that case there may still be efficiency gains in combining the data, as discussed in \citet{rosenman2020combining}. However, in large samples the two estimators will be converging to different limits. 
  There is no information in the data to determine whether unconfoundedness in the observational sample, or external validity in the experimental sample, is violated.  Note that these assumptions  are of a very different nature. The researcher has to use {\it a priori} arguments to choose between $\hat\tau^{\pps,\ppe}$ and $\hat\tau^{\pps,\ppo}$ and the corresponding assumptions.
Choosing for $\hat\tau^{\pps,\ppe}$ would imply being less concerned with the external validity of the experimental sample, whereas the choice for $\hat\tau^{\pps,\ppo}$ would imply that the internal validity of the observational study would be viewed as less of a concern.
In many cases researchers have argued for the primacy of internal validity over external validity ({\it e.g.}, \citet{shadishcookcampbell, imbens2010}), though others have argued against that perspective ({\it e.g.}, \citet{manski2013public, deaton2010}).
If one prefers $\hat\tau^{\pps,\ppo}$ (downplaying the concerns about internal validity of the observational sample), the natural estimator for $\tau^\pp$ is $\hat\tau^{\pp,\ppo}$, and there is little use for the experimental sample. However, if the researcher prefers 
$\hat\tau^{\pps,\ppe}$, the question arises how to estimate $\tau^\pp$.

 The current paper is concerned with this question. We take the position that there is  an {\it a priori} preference for  $\hat\tau^{\pps,\ppe}$ over $\hat\tau^{\pps,\ppo}$, possibly after accounting for differences in covariate distributions to deal with some of the external validity concerns (\citet{hotz2005predicting}). Then we address the  main question of how to adjust the estimator $\hat\tau^{\pp,\ppo}$ to take into account the difference between $\hat\tau^{\pps,\ppo}$ and $\hat\tau^{\pps,\ppe}$,
Conceptually there are multiple natural ways of doing so. We discuss three of these ways. The first  one is based on imputation of the missing primary outcomes in the experimental sample. The second one is  based on weighting the units in the observational sample. The third one is based on a control function approach.
In this simple nonparametric case with binary outcomes the three approaches lead to identical point estimates .

\subsubsection{Imputation}

The first approach is to take a missing data perspective on the primary $Y_i^\pp$ in the experimental sample and  impute these missing values using the observational 
sample. Consider  unit $i$ in the experimental sample with $W_i=w$ and $Y^\pps_i=y^\pps$. Assuming the missing data on $Y^\pp_i$  are missing at random ({\it e.g.,} \citet{rubin1976inference, little2019statistical, rubin2004multiple}) suggests using the distribution of $Y^\pp_i$ among units in the observational sample with $W_i=w$ and $Y^\pps_i=y^\pps$ to impute the missing values. If we are interested in estimating the average effect, we can just use the average value of $Y^\pp_i$ in this subsample as the imputed value.
Denote  this average value for all values of $y^\pps$ and $w$ by: 
\[ \oy^{\pp,\ppo}_{w,y^\pps}=
\sum_{i=1}^N \been_{G_i=\ppo, W_i=w,Y^\pps_i=y^\pps}Y^\pp_i
\Bigl/\sum_{i=1}^N \been_{G_i=\ppo, W_i=w,Y^\pps_i=y^\pps}.\]
The imputed value for $Y^\pp_i$ for  unit $i$ in the experimental sample is then the average of $Y^\pp_j$ in the 
observational sample over all units $j$  with the same treatment level, $W_j=W_i$, and the same value for the secondary outcome, $Y^\pps_j=Y^\pps_i$:
\[ \hat Y^\pp_i=\oy^{\pp,\ppo}_{W_i,Y^\pps_i}.\]
Then the imputation estimator for $\tau^\pp$ is the difference in average imputed values in the experimental sample by treatment status, leading to the first estimator for $\tau^\pp$:
\begin{equation}\label{imp}\hat\tau^{\pp,\imp}\equiv \frac{1}{N^\ppe_1}
\sum_{i:G_i=\ppe} W_i \oy^{\pp,\ppo}_{W_i,Y^\pps_i}
-\frac{1}{N^\ppe_0}\sum_{i:G_i=\ppe} (1-W_i) \oy^{\pp,\ppo}_{W_i,Y^\pps_i}
.\end{equation}

\subsubsection{Weighting}

The second approach to using the experimental secondary outcomes  is to reweight the observational sample where we do observe the primary outcome. Consider the $N^\ppo_w$ units in the observational sample with $W_i=w$. The fraction of those treated units  in the observational study with secondary outcome $Y^\pps_i=1$ is $\hat p^{\pps,\ppo}_w=\sum_{i:G_i=\ppo,W_i=w} Y^\pps_i/\sum_{i:G_i=\ppo,W_i=w} 1$. The experimental study tells us this fraction would have been approximately
$\hat p^{\pps,\ppe}_w=\sum_{i:G_i=\ppe,W_i=w} Y^\pps_i/\sum_{i:G_i=\ppe,W_i=w} 1$, had the treatment  been randomly assigned. 
The comparison $\hat p^{\pps,\ppo}_w$ versus $\hat p^{\pps,\ppe}_w$ reflects on the possible violation of the unconfoundedness assumption in the observational sample.
We can give these units a weight $\lambda_{w,1}=\hat p^{\pps,\ppe}_w/\hat p^{\pps,\ppo}_w$ to adjust for the bias stemming from such violations.
Similarly, units with treatment $W_i=w$ and $Y^\pps_i=0$ would be given a weight $\lambda_{w,0}=(1-\hat p^{\pps,\ppe}_w)/(1-\hat p^{\pps,\ppo}_w)$. We then
 use these to estimate the average effect on the primary outcome as
the difference of weighted averages of the treated and control outcomes in the observational sample, leading to the second estimator:
\begin{equation}\hat\tau^{\pp,\weight}\equiv 
\frac{\sum_{G_i=\ppo} \lambda_{1,Y^\pps_i} W_i Y^\pp_i}{\sum_{G_i=\ppo} \lambda_{1,Y^\pps_i}W_i }
-\frac{\sum_{G_i=\ppo} \lambda_{0,Y^\pps_i}(1-W_i) Y^\pp_i}{\sum_{G_i=\ppo} \lambda_{0,Y^\pps_i} (1-W_i)}
.\end{equation}
Simple algebra shows that  the two estimators for $\tau$ are identical,
$\hat\tau^{\pp,\imp}=\hat\tau^{\pp,\weight}$.
This algebraic result relies on the outcome model and the weights being fully nonparametric in this simple example with the secondary outcome taking on only two values. In settings with the secondary outcome continuous, the two approaches will generally give different answers in finite samples.

\subsection{A Control Function Approach in a  Linear Model Setting}

In our second example 
 we consider a simple linear model that exhibits most clearly some of the key features of the approach in the current paper. 
Specifically, suppose we have a linear model for the secondary potential outcomes with a constant treatment effect:
\[ Y^{\pps}_i(0)=X_i^\top\gamma^\pps+\alpha_i^\pps,\hskip1cm Y^\pps_i(1)=Y^{\pps}_i(0)+\tau^\pps
.\]
This models holds for both the experimental and observational samples.
The properties of the  unobserved component $\alpha^\pps_i$ are key, and they may differ in the two samples. In the experimental sample the randomization guarantees that we have the following conditional independence: 
\[ W_i\  \perp\!\!\!\perp\ \alpha_i^\pps\ \Bigl|\ X_i,G_i=\ppe.\]
In fact the randomization implies even stronger conditions, 
$W_i \perp\!\!\!\perp\alpha_i^\pps,X_i|G_i=\ppe$,
but we do not need those here.
In the observational study we do not in general have the same conditional independence:
\[ W_i\  \not\!\perp\!\!\!\perp\ \alpha_i^\pps\ \Bigl|\ X_i,G_i=\ppo.\]
 This randomization in the experimental sample implies that we can estimate the parameters of the model for the secondary outcome,  $\tau^\pps$ and $\gamma^\pps$,  by least squares regression of $Y^\pps_i$ on $W_i$ and $X_i$ using only the data from the experimental sample.
 In other words, the conditional mean of $Y^\pps_i$ given $W_i$ and $X_i$ has a causal interpretation as a function of $W_i$ in the experimental sample, but not in the observational sample.
 
Now consider the primary outcome.
We specify a similar linear
 model for the primary outcome, but allow the coefficients to be different from those of the model for the secondary outcome,
\[ Y^{\pp}_i(0)=X_i^\top\gamma^\pp+\alpha_i^\pp,\hskip1cm  Y^{\pp}_i(1)= Y^{\pp}_i(0)+\tau^\pp
.\]
Again the concern is that in the observational sample the unobserved component might be correlated with the treatment:
\[W_i\  \not\!\perp\!\!\!\perp\ \alpha_i^\pp\ \Bigl|\ X_i,G_i=\ppo.\]
Such a correlation would imply that a linear regression of $Y^\pp_i$ on $W_i$ and $X_i$ using the data from the observational sample would not be consistent for the causal effect $\tau^\pp$ because of endogeneity of $W_i$.
Now a key assumption is that there is a relationship between the short term and long term unobserved components $\alpha^\pp_i$ and $\alpha^\pp_i$ that allows us to remove the endogeneity bias in the long term relationship using the difference between the short term results for the experimental and observational data using a control function approach (\citet{heckman1985alternative, imbens2009identification, kline2019heckits}). The key assumption that links the endogeneity problems for the primary and secondary outcomes is
\begin{equation}\label{cf_linear}\alpha^\pp_i=\delta\alpha^\pps_i+\varepsilon_i^\pp,
\hskip1cm {\rm with}\ \  
W_i\  \perp\!\!\!\perp\ \varepsilon_i^\pp\ \Bigl|\ X_i,\alpha^\pps_i,G_i=\ppo.\end{equation}
Later we relax this assumption to remove the functional form dependence, but for the moment let us focus on this version with linearity and additivity. The key is that the residual for the primary outcome, $\alpha^\pp_i$, is related to the residual for the secondary outcome, $\alpha^\pps_i$, with the remainder, $\varepsilon_i^\pp=\alpha_i^\pp-\mme[\alpha^\pp_i|\alpha^\pps_i]$  unrelated to the treatment.

Let us show in some detail how this assumptions aids in the identification of $\tau^\pp$ in this linear example.
First,  we can estimate $\tau^\pps$ and $\gamma^\pps$ from the experimental sample by linear regression. Denote these least squares estimates by $\hat\tau^{\pps}$ and $\hat\gamma^{\pps}$. Then  we can estimate the residual $\alpha_i^\pps$ for the units in the observational sample as
\begin{equation}\label{alpha} \hat \alpha^\pps_i=Y^\pps_i-W_i\hat\tau^{\pps}-X_i^\top\hat\gamma^{\pps}.\end{equation}
If this model is correct, and if the assignment to treatment in the observational sample were random, and finally, if  the observational and experimental samples were randomly drawn from the same population, the population value of these residuals $\alpha^\pps_i$ would  have mean zero and be uncorrelated with the treatment indicator in the observational sample. The presence of non-zero association of this residuals and  the treatment is exploited to adjust the estimates of the treatment effect on the primary outcome.
We can do so by including this residual as a control variable in the least squares regression with the long term outcome as the dependent variable, using the observational data. The key insight is that we can
use the linear representation in (\ref{cf_linear}) to write the long te outcome as:
\begin{equation}\label{control} Y^{\pp}_i=W_i\tau+X_i^\top\gamma+\delta\alpha^\pps_i+\varepsilon^\pp_i,\hskip1cm
{\rm with}\ \ 
W_i  \perp\!\!\!\perp\ \varepsilon_i^\pp\ \Bigl|\ X_i,\alpha^\pps_i,G_i=\ppo.\end{equation}
Therefore this regression will lead to a consistent estimator for $\tau^\pp$ under the current assumptions.

To further develop intuition for the control function approach, consider the example where  the primary and secondary outcomes are eight and third grade test scores, and the treatment is class size. Using the experimental sample we estimate the the average effect of the class size on third grade scores. We then calculate the residuals in the observational study. We may find that the residuals are larger on average for the treated individuals than for the control individuals. This suggests that the treatment assignment in the observational sample was correlated with the third grade potential outcomes, with individuals with high values for the potential outcomes more likely to be in the treatment group. We then use that information to compare eighth grade scores for individuals with the same residuals, so we adjust for the original non-random selection into treatment.

\subsection{The Connection Between the Imputation and Control Function Approaches}

The control function approach to dealing with the endogeneity of the treatment in the observational study we used in the second example may appear at first sight to be conceptually quite different from the weighting and imputation approaches in the first example. In fact the two approaches are closely related. Consider the imputation of $Y^\pp_i$ for a unit in the experimental sample given the linear model. 
Substituting for $\alpha^\pps_i$ using Equation (\ref{alpha}) into Equation (\ref{control}) implies that we can write for the primary outcome in the observational sample:
\[ Y^{\pp}_i=W_i\beta+X_i^\top\lambda+\delta Y^\pps_i+\varepsilon_i,\]
where
\[ \beta=\tau^\pp-\delta\tau^\pps,\hskip1cm{\rm and}\ \  \lambda=\gamma^\pp-\delta\gamma^\pps.\]
Hence the imputed value  for $Y^\pp_i$ in the experimental sample using the estimated parameters from the observational sample leads to (ignoring estimating error)
\[ \hat  Y^{\pp}_i=W_i\beta+X_i^\top\lambda+\delta Y^\pps_i.\]
Using the omitted variable bias formula it is easy to see that regressing this imputed value $\hat Y^\pp_i$ on $W_i$ and $X_i$ (but omitting $Y^\pps_i$), in the experimental sample, leads to
\[ \hat Y^\pp_i=W_i\bar \beta+X_i^\top\bar \lambda+\varepsilon_i,\hskip1cm {\rm
with}\ \ 
 \bar\beta=\beta+\delta\beta^\pps=\tau^\pp.\]
Thus, the coefficient on $W_i$ in this regression of the imputed primary outcome on the treatment and the pretreatment variables is consistent for the causal effect of the treatment on the primary outcome.

\section{The General Case}

The two examples in the preceeding section convey much of the intuition for our approach:. 
The key assumption in the linear case is (\ref{cf_linear}), which connects the bias in causal estimates for the primary and secondary outcomesin the observational sample.
Making such assumptions allows us  improve upon estimates for $\tau^\pp$  based on the observational sample alone. What we do in this section is generalize the first example to the case where $(i)$ the secondary and primary outcomes may be continuous and $(ii)$ the secondary outcome may be vector-valued, and $(iii)$ where pre-treatment variables are present. We also generalize the control function approach to $(i)$ the nonlinear case, and $(ii)$ the case  with multiple secondary outcomes in order to allow the critical assumptions to be weakened.
We present the formal assumptions that justify the weighting and imputation estimators, and present their general forms and how they relate to the control function approach.



We are interested in causal estimands defined for the population of interest. At a general level such estimands include simple average treatment effects, but more generally also the average effect of a policy that assigns the treatment to individuals in this population on the basis of covariates ({\it e.g.}, \citet{manski2004statistical, dehejia2005program, hirano2009asymptotics, athey2017efficient, zhou2018offline}). 

Define
\begin{equation}\label{tau_general}\tau_{g}^{t}\equiv \mme\left[\left.Y_{i}^{t}(1)-Y_{i}^{t}(0)\right|G_{i}=g\right],\end{equation}
is the average effect of the treatment on outcome $t\in\{\pps,\pp\}$ for group $g\in\{\ppo,\ppe\}$.
The superscripts on the estimands denote the outcome, and subscripts denote the population.
The  primary estimand we focus on in this paper   is the average effect of the treatment on the long term outcome in the observational study population: 
\begin{equation}\label{tau}\tau\equiv \tau_{\ppo}^{\pp}\equiv \mme\left[\left.Y_{i}^{\pp}(1)-Y_{i}^{\pp}(0)\right|G_{i}=\ppo\right],\end{equation}
where we drop the subscript and superscript to simplify the notation.

\subsection{Three Maintained Assumptions}

There are three key features of our set up. First, we are interested in the population that the units in the observational study were drawn from. That is, the observational study has external validity.
\begin{assumption}{\sc (External Validity of the Observational Study)} The observational sample is a random sample of the population of interest.
\label{assumption:external_observational}
\end{assumption} At some level this can be thought of as simply defining the estimand in terms of the population distribution underlying the observational sample.

Second, we maintain throughout the paper the assumption that 
the treatment in the  experimental sample is unconfounded.
\begin{assumption}\label{assumption:random}{\sc (Internal Validity of the Experimental Sample)}
For $w=0,1$,
\begin{equation}\label{random} W_i\ \perp\!\!\!\perp\ \Bigl(Y_{i}^{\pp}(w),Y_{i}^{\pps}(w)\Bigr)\ \Bigr| \ X_i,G_i=\ppe.\end{equation}
\label{assumption:internal_experimental}\end{assumption}
\citet{kallus2020role}  make a different assumption here,
\begin{equation}\label{kallus}W_i\ \perp\!\!\!\perp\ \Bigl(Y_{i}^{\pp}(w),Y_{i}^{\pps}(w)\Bigr)\ \Bigr| \ X_i,\end{equation}
where unconfoundedness holds in the combined sample, rather than in the experimental sample.
Assumption (\ref{kallus}) does not imply our assumption (\ref{random}), or the other way around. In our application, with assignment in the Project STAR experimental sample completely randomized, our assumption is satisfied by design, and in general (\ref{kallus})  would not hold. In other settings, for example where the data are sampled from a single population rather than two separate populations,  (\ref{kallus})  may be more appropriate than our assumption.

 However, the external validity of the experimental study is not guaranteed.
Instead we assume that conditional on the pretreatment variables we have external validity (\citet{hotz2005predicting}):
\begin{assumption}\label{assumption:conditional}{\sc (Conditional External Validity)} The experimental study has  conditional external validity if
\begin{equation}\label{eq:cond_ext} G_i\ \perp\!\!\!\perp\ \Bigl(Y_{i}^{\pp}(0),Y_{i}^{\pp}(1),
Y_{i}^{\pps}(0),Y_{i}^{\pps}(1)\Bigr)\ \Bigr|\ X_i.\end{equation}
\end{assumption}
This assumption implies that if we find systematic differences between in differences in average outcomes by treatment status conditional on covariates between the experimental and observational sample, these differences must arise from violations of unconfoundedness for the observational sample.



The first result is that these three maintained assumptions are in general not sufficient for point-identification of the average effect of interest. Of course this does not mean that these assumptions do not have any identifying power. They do in fact affect the identified sets in the spirit of the work by (\citet{manski_bounds}).

\begin{lemma}\label{lemma1}
The combination of  Assumptions \ref{assumption:external_observational}-\ref{assumption:conditional} is not sufficient for point-identification of $\tau^\pp$.
\end{lemma}
The proof for this result is given in the appendix.

\subsection{Unconfoundedness for the Observational Sample}

Next, let us consider the assumption that assignment in the observational study is unconfounded.
\begin{assumption}\label{assumption:unconfoundedness}
{\sc (Unconfoundedness in the Observational Sample)}\\
 For $w=0,1$,
\begin{equation}W_{i}\ \perp\!\!\!\perp\ \Bigl(Y_{i}^{\pps}(w),Y_{i}^{\pp}(w)\Bigr)\ \Bigl|\ X_{i},G_i=\ppo,\end{equation}
\end{assumption}
This assumption is made, for example, in \citet{rosenman2018propensity}. 
This assumption is sufficient for identification of $\tau$, but it is stronger than necessary. Intuitively it implies that we do not need the experimental sample for identification because under unconfoundedness the observational sample is sufficient for identification of the average treatment effect. However, the experimental sample may still be useful for precision.
The precise version of the unconfoundedness assumption here is slightly different from that in, say, \citet{rosenbaum1983central} where it is assumed that $W_i$ is independent of the full set of $(Y_i^\pp(w),Y^\pps_i(w))_{w\in\{0,1\}}$. It is what is referred to in \citet{imbens2000} as ``weak unconfoundedness.'' This issue will come up later.

\begin{lemma}\label{lemma2}
The set  of  Assumptions \ref{assumption:random}-\ref{assumption:unconfoundedness} has a testable implication:
\begin{equation} \label{eq:unconf2} G_i\ \perp\!\!\!\perp\ Y^\pps_i\ \Bigl|\ X_i,W_i.\end{equation}
\end{lemma}

We can also use this result to assess whether a particular set of pre-treatment variables is sufficient for unconfoundedness. Here we are interested in finding a set of pretreatment variables $X_i$ such that
\begin{equation}
G_i\ \perp Y^\pps_i\ \Bigl|\ X_i=x,W_i,
\end{equation}
holds.

\subsection{Latent Unconfoundedness}

Suppose that we reject the  conditional independence in Lemma \ref{lemma2}, so that we know that the full set of maintained assumptions, \ref{assumption:random}-\ref{assumption:unconfoundedness}, does not hold. If we maintain unconfoundedness in the experimental sample, it must be that either  conditional external validity in the experimental study, or unconfoundedness in the observational study must be violated.
If we interpret such a finding as evidence against conditional external validity, and are willing to maintain unconfoundedness of the treatment assignment in the observational study, we should simply put aside the experimental data set and focus on estimates based on solely on the observational study. In many cases, however, we may wish to maintain conditional external validity and  interpret  a finding that 
(\ref{eq:unconf2})  does not hold as evidence that unconfoundedness does not hold for the observational study. Here we explore methods for using the difference between the estimates of the causal effects for the experimental study (which we know to be internally valid) and the estimates for the observational study (which need not be internally valid) to adjust long term estimates for the observational study.

The idea, although not the implementation, is somewhat similar to that in a Difference-In-Differences  (\citet{cardmariel, cardkrueger1, angrist2008mostly}) set up where the initial (pre-treatment) differences between a treatment and control group are used to adjust post-treatment differences between the treatment and control group. More specifically, it relates to the Changes-In-Changes approach in \citet{athey2006identification} where functional form assumptions are avoided. Here initial differences in treatment effects between an experimental and observational study are used to adjust subsequent treatment effects for the observational study.

They key additional assumption that links the biases, in the observational study, between adjusted comparisons for the primary and secondary outcomes, is the following.
\begin{assumption}\label{assumption:new}{\sc (Latent Unconfoundedness)}\\
For $w\in\{0,1\}$,
\begin{equation}\label{new}
W_i\ \perp\!\!\!\perp\ Y_{i}^{\pp}(w)\ \Bigl|\ X_{i},Y^\pps_i(w),G_i=\ppo
.\end{equation}
\end{assumption}
This assumption is both novel as well as  critical in the current discussion, so let us offer some remarks.
\begin{remark}
Compared to a regular unconfoundedness assumption, we add the variable $Y^\pps_i(w)$ to the conditioning set. At first this may appear to be an innocuous addition. However, following the standard approach to exploiting unconfoundedness assumptions, we see that this is not the case. Typically we use such an assumption to create subpopulations defined by the conditioning variables, and then compare treated and control units.  To be specific, suppose we wish to estimate $\mme[Y^\pp_i(1)|G_i=\ppo]$. We would first estimate the conditional expectation 
 $\mme[Y^\pp_i(1)|Y_i^\pps(1)=y^\pps,W_i=1,X_i=x,G_i=\ppo]$. Then, however, we would need to average this over the marginal distribution of $(Y_i^\pps(1),X_i)$ in the observational sample, but in this observational sample we only see draws from the conditional distribution of $(Y_i^\pps(1),X_i)$ given $W_i=1$, and this is not the same distribution because of the failure of unconfoundedness in the observational sample. To address this, we need to exploit the presence of the experimental sample.
\end{remark}

To highlight the link to the control function literature (\citet{heckman1979sample, heckman1985alternative, imbens2009identification, wooldridge2010econometric, athey2006identification, kline2019heckits, mogstad2018using, mogstad2018identification, wooldridge2015control}), let us model the primary and secondary   outcomes as
\[Y^\pp_i(w)=h^\pp(w,\nu_i,X_i),\hskip1cm {\rm and}\ \   Y^\pps_i(w)=h^\pps(w,\eta_i,X_i)
,\]
with $h^\pps(w,\eta,x)$ strictly monotone in $\eta$.
Now we can write the latent unconfoundedness assumption as
\[ W_i\ \perp\!\!\!\perp\ \nu_i\ \Bigl|\ X_{i},\eta_i,G_i=\ppo.\]
Although it is not generally true that $W_i \perp\!\!\!\perp \nu_i| X_{i},G_i=\ppo$, adding $\eta_i$ to the conditioning set restores the exogeneity of $W_i$ in the observational sample.

It is useful to contrast this with a control function in a non-parametric instrumental variables setting ({\it e.g.,} \citet{imbens2009identification}), where the two models  are
\[Y^\pp_i(w)=h^\pp(w,\nu_i,X_i),\hskip1cm {\rm and}\ \   W_i(z)=r(z,\eta_i,X_i)
 ,\]
 with $r(z,\eta,x)$ strictly monotone in $\eta$.
The key assumption here is that
\[  W_i\ \perp\!\!\!\perp\ \nu_i\ \Bigl|\ X_{i},\eta_i.\]
The model relating the outcome of interest and the endogenous regressor is essentially the same in the two settings, $Y^\pp_i(w)=h^\pp(w,\nu_i,X_i)$. In both cases we address the endogeneity by conditioning on an additional variable, the control variable $\eta_i$. This control variable is estimated using an auxiliary model. This auxilliary model  differs between the set up in the current paper and the instrumental variables setting.
In the instrumental variables setting we  model the relation between the endogenous regressor and an additional variable, the instrument, and deriving the control variable from that relation.
In the current setting we  model the relation between the secondary outcome and the endogenous regressor and deriving the control variable from that relation. In both cases the auxiliary model has a strict monotonicity assumption. This shows some of the limitations of the approach: the unobserved confounder $\eta$ cannot have a dimension higher than that of the secondary outcome.

Formally, adding  Assumption \ref{assumption:new} (latent unconfoundedness) to Assumptions \ref{assumption:external_observational}-\ref{assumption:conditional} allows us to point-identify the average effect of interest.
\begin{theorem}\label{theorem_main}
Suppose that Assumptions \ref{assumption:external_observational}-\ref{assumption:conditional} and \ref{assumption:new} hold, so  that the experimental study is unconfounded and has conditional external validity, and the observational study has latent unconfoundedness. Then 
 $\tau^\pp_\ppo$,
the average effect of the treatment on the primary outcome in the observational study is point-identified.
\end{theorem}
\subsection{Missing At Random}

There is an interesting and  close connection between Assumptions \ref{assumption:external_observational}-\ref{assumption:conditional} and \ref{assumption:new}  and the Missing-At-Random (MAR) assumption in the missing data literature (\citet{rubin1976inference, little2019statistical, rubin2004multiple}).
\begin{lemma}
Suppose  that Assumptions \ref{assumption:external_observational}-\ref{assumption:conditional} and \ref{assumption:new}  hold. Then:
\begin{equation}\label{mar}G_i\ \perp\!\!\!\perp\ Y_{i}^{\pp}\ \Bigl|\ W_{i},X_{i},Y_{i}^{\pps}.\end{equation}
\end{lemma}
Because $G_i=\ppe$ is equivalent to an indicator that $Y^\pp_i$ missing, and because $W_i$, $X_i$, and $Y^\pps_i$ are observed for all individuals in the sample, the conditional independence in (\ref{mar}) is equivalent to a MAR assumption. The result does not go the other way around. The MAR assumption by itself has no testable implications, but the combination of Assumptions 
Assumptions \ref{assumption:external_observational}-\ref{assumption:conditional} and \ref{assumption:new} does imply some inequality restrictions on the joint distribution of the observed variables.
\citet{kallus2020role} starts with a Missing-At-Random assumption, and uses that in combination with an unconfoundedness assumption on the full sample to identify the average effect of the treatment for the full sample.

\section{Estimation and Inference}

In this section we discuss formal nonparametric estimation strategies, some of which we discussed in the examples in Section \ref{two_examples}. These strategies include imputation, weighting, control function methods, and influence-function based methods for the general case. 
\citet{chen2018overidentification}

\subsection{The Imputation Approach}
First, consider the imputation approach. 
Estimate the conditional mean of the primary outcome given the secondary outcome, treatment and pre-treatment variables in the observational sample:
\[\kappa(w,x,y,\ppo)\equiv \mme\left[ \left.Y_i^\pp\right| W_i=w,X_i=x,Y^\pps_i=y,G_i=\ppo\right].\]
Then impute for the units in the experimental sample the primary outcome as $\hat Y^\pp_i=\hat{\kappa}(W_i,X_i,Y^\pps_i,\ppo)$.
Then use the standard program evaluation methods to adjust for differences in covariates if necessary. If in the experimental sample the treatment is completely random, we would estimate the average treatmet effect in the experimental sample as
\[ \hat\tau^{{\rm imp},\ppe}=\frac{1}{N^\ppe_1}\sum_{i:P_i=\ppe} W_i \kappa(1,X_i,Y^\pps_i,\ppo)-
\frac{1}{N^\ppe_1}\sum_{i:P_i=\ppe} (1-W_i)\kappa(0,X_i,Y^\pps_i,\ppo).\]
However, we wish to estimate the average effect in the observational sample, which may have a different distribution of the pre-treatment variables. This requires one additional layer of adjustment that depends on the pre-treatment variables. Define
\[ r(x)={\rm pr}(G_i=\ppo|X_i=x).\]
Then we weight the units by the ratio $r(X_i)/(1-r(X_i))$:
\[ \hat\tau^{{\rm imp}}=\frac{\sum_{i:P_i=\ppe} W_i \kappa(1,X_i,Y^\pps_i,\ppo) r(X_i)/(1-r(X_i)}{\sum_{i:P_i=\ppe} W_i r(X_i)/(1-r(X_i)}\]
\[\hskip2cm -
\frac{\sum_{i:P_i=\ppe} (1-W_i)\kappa(0,X_i,Y^\pps_i,\ppo)r(X_i)/(1-r(X_i))}{\sum_{i:P_i=\ppe} (1-W_i)r(X_i)/(1-r(X_i))}.\]

\subsection{The Weighting Approach}
Second, consider the weighting approach. Estimate the distribution of $(Y^\pps_i,W_i)$ in the observational and experimental sample as
\[ f_{W,Y^\pps|X,P}(w,y^\pps|x,p),\]
for all $x\in\mmx$ and $p\in\{\ppe,\ppo\}$.
Then  construct the weights for all units in the observational sample as a function of $(W_i,X_i,Y^\pps_i)$:
\[ \lambda_i=\frac{f_{W,Y^\pps|X,P}(W_i,Y_i^\pps|X_i,\ppe)}{f_{W,Y^\pps|X,P}(W_i,Y_i^\pps|X_i,\ppo)}.\]
These weights adjust for the differences between the observational and experimental sample.

Assuming we have completely random assignment in the experimental sample,  we estimate the average treatment effect as
\[ \hat\tau^{\rm weight}=\frac{\sum_{i:P_i=\ppo} Y_i W_i \lambda_i}
{\sum_{i:P_i=\ppo}  (1-W_i) \lambda_i}-
\frac{\sum_{i:P_i=\ppo} (1-W_i) \lambda_i}
{\sum_{i:P_i=\ppo}  W_i \lambda_i}.\]

If instead we have unconfounded treatment assignment in the experimental sample, we need the weights that adjust for the non-randomness in the experimental sample. By the maintained assumptions this requires only adjusting for the differences in pre-treatment variables. Let the propensity score be
\[ e(x,g)={\rm pr}(W_i=1|X_i=x,G_i=g).\]
This leads to
\[ \hat\tau^{\rm weight}=\frac{\sum_{i:P_i=\ppo} Y_i W_i \lambda_i/e(X_i,\ppe)}
{\sum_{i:P_i=\ppo}  (1-W_i) \lambda_i/e(X_i,\ppe)}-
\frac{\sum_{i:P_i=\ppo} (1-W_i) \lambda_i/(1-e(X_i,\ppe))}
{\sum_{i:P_i=\ppo}  W_i \lambda_i/(1-e(X_i,\ppe))}.\]

\subsection{The Control Function Approach}
Finally, the control function approach. First estimate the conditional distribution of the secondary outcome given  treatment and pre-treatment variables in both samples:
\[ F_{Y^\pps|W,X,G}(y^\pps|w,x,g)={\rm pr}(Y^\pps_i\leq y^\pps|W_i=w,X_i=x,G_i=g).\]
Then calculate the control variable for each unit in the observational sample as
\[ \hat\eta_i=\hat F_{Y^\pps|W,X,G}(Y_i^\pps|W_i,X_i,\ppe).\]
Next, estimate the conditional mean of the primary outcome in the observational sample given treatment status, control variable, and pre-treatment variables:
\[\gamma(w,h,x)\equiv  \mme\left[\left. Y^\pp_i\right|W_i=w, \eta_i=h,X_i=x,G_i=\ppo\right].\]
Finally, estimate the average treatment effect $\tau$ as
\[ \hat\tau^{\rm cf}=\frac{1}{N^\ppe_1}\sum_{i:G_i=\ppe} W_i \hat\gamma(1,\hat\eta_i,X_i)-
\frac{1}{N^\ppe_0}\sum_{i:G_i=\ppe} (1-W_i) \hat\gamma(1,\hat\eta_i,X_i).\]

\section{An Application}

To illustrate the ideas in this paper we analyze data on the effect of class size on educational outcomes. We use the data from the Project STAR experiment on class size, where we observe  for all children whether they are in a regular or small class. As the short term outcome we use the third grade score. We also observe the pre-treatment variables gender, whether the student gets a free lunch, and ethnicity. For the observational data we use data from the New York school system. We observe the same variables, but also eighth grade scores.

In Table \ref{tab:} we report the results from several ordinary least squares regressions. The first two columns show the results from a regression of the short term outcome on the treatment, separately for the two samples, and controlling for the pretreatment variables.
 In the experimental sample in Project STAR we find a positive effect of small class size of 0.157. In the observational sample the least squares estimate is negative, -0.048. This suggests that there are unmeasured confounders. If we regress the eighth grade scores on the treatment in the observational sample we still get a negative estimate, -0.074. 
Now we follow the control function approach and include in that linear regression the control function, that is, the estimated residual:
\[ \hat \alpha^\pps_i=Y^\pps_i-W_i\hat\beta^\pps-X_i^\top\hat\gamma^\pps_i.\]
Including this in the regression gives a coefficient of 0.640 with a standard error of 0.001. It changes the coefficient on the treatment to 0.061, now much more in line with what one would expect given the causal effect of a small class size on the third grade scores in the experimental sample.
We can also do this through the imputation approach. First we use the regression of the long term outcome on short term outcome and covariates to predict the long term outcome for the observations in the experimental sample. Then we regress the predicted value on the treatment, leading to the same estimate of 0.061.


\begin{table}[tbh]
        \caption{}
        \label{tab:}
        \centering
        \vspace{1em}
        \begin{tabular}{lccccc}
\toprule
{} &                                      Secondary &                                     Secondary  &                                        Primary &                                       Primary  &                                Imputed Primary \\
\midrule\\
$W_i$            &                      $0.157$\textsuperscript{} &                     $-0.048$\textsuperscript{} &                     $-0.074$\textsuperscript{} &                      $0.061$\textsuperscript{} &                      $0.061$\textsuperscript{} \\
                 &  ($0.028$\textsuperscript{})\textsuperscript{} &  ($0.002$\textsuperscript{})\textsuperscript{} &  ($0.003$\textsuperscript{})\textsuperscript{} &  ($0.018$\textsuperscript{})\textsuperscript{} &  ($0.018$\textsuperscript{})\textsuperscript{} \\ \\
$\hat\alpha_i^S$ &                                                &                                                &                                                &                      $0.640$\textsuperscript{} &                                                \\ 
                 &                                                &                                                &                                                &  ($0.001$\textsuperscript{})\textsuperscript{} &                                                \\ \\
$N$              &                      $6,027$\textsuperscript{} &                  $1,131,339$\textsuperscript{} &                    $498,597$\textsuperscript{} &                    $498,597$\textsuperscript{} &                      $6,027$\textsuperscript{} \\
$R^2$            &                      $0.130$\textsuperscript{} &                      $0.060$\textsuperscript{} &                      $0.040$\textsuperscript{} &                      $0.420$\textsuperscript{} &                      $0.170$\textsuperscript{} \\
Sample           &                                   Experimental &                                  Observational &                                  Observational &                                  Observational &                                   Experimental \\
Covariates       &                                            Yes &                                            Yes &                                            Yes &                                            Yes &                                            Yes \\
\bottomrule
\end{tabular}
        \end{table}

    \begin{figure}[tb]
        \centering
        \includegraphics[]{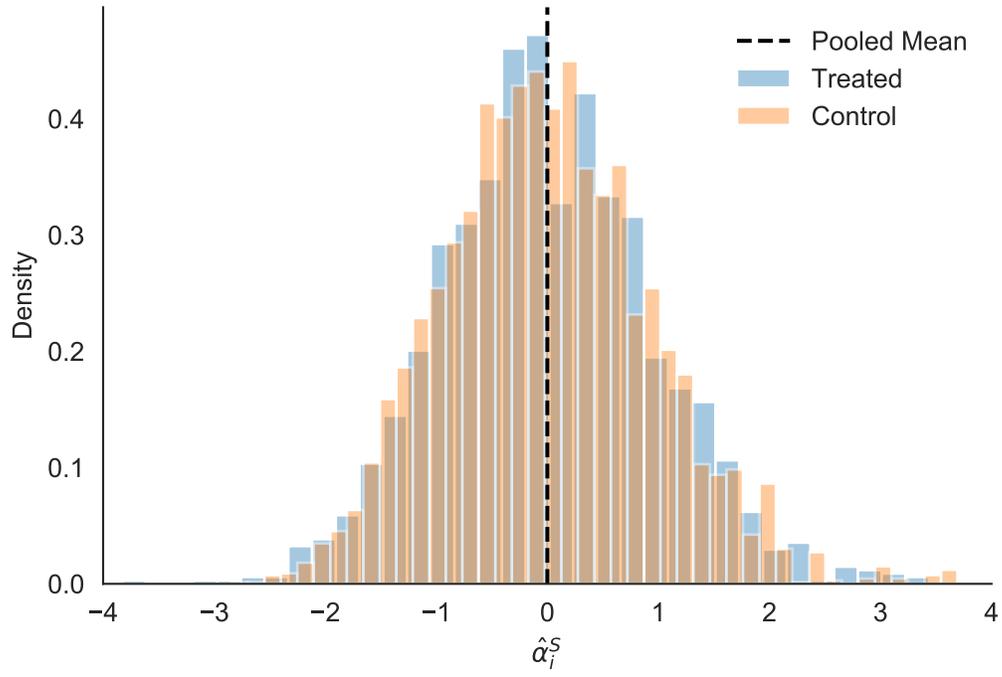}
        \caption{$\alpha^S_i$ in the experimental sample by treated and
        control}
        \label{fig:figure1}
    \end{figure}
        \begin{figure}[tb]
        \centering
        \includegraphics[]{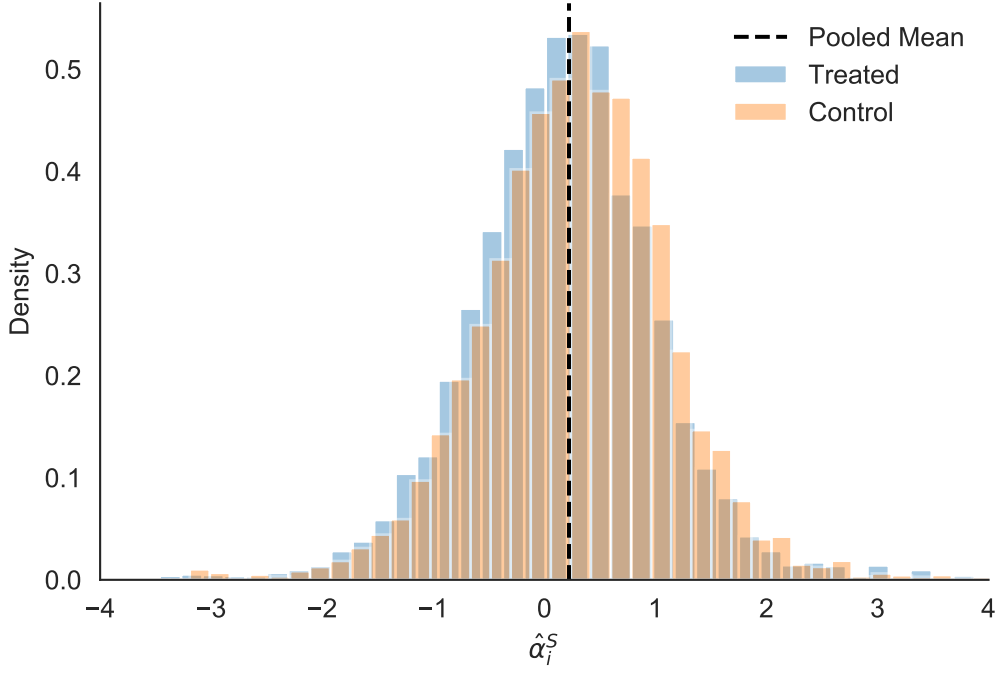}
        \caption{$\alpha^S_i$ in the observational sample by treated and
        control}
        \label{fig:figure2}
    \end{figure}

 We also investigate if the surrogacy assumption holds here. Estimating a regression of the primary outcome on the secondary outcome and the treatment indicator (and including pre-treatment variables), leads to a coefficient on the treatment indicator of -0.039, with a standard error of 0.002. Thus, it appears that third grade scores are not a valid surrogate for eighth grade scores.  Thus, the effect of class size is not fully captured by third grade scores, but also arises through other channels. Thus, accounting for the latent confounder is important.

\section{Conclusion}

In this paper, we develop new statistical methods for systematically combining experimental and observational data in an attempt to leverage the internal validity of the experimental studies and the external validity and high precision of the observational studies. We 
do so in a setting where the experimental sample contains information on a secondary outcome and the observational study contains information on both primary and secondary outcomes.
We articular a new and critical assumption that allows us to link the biases in comparisons in the observational study between primary and secondary outcome exploing the bias-free information on the secondary outcome from the experimental data. We illustrate these new results by 
 combining data from the  Project STAR experiment with observational data from the New York school system. We find that the biases in the observational study are substantial, but that the adjustment procedure based on the experimental data leads to more plausible results.

\newpage

\citet{adams2006overweight, d2006surrogate, abadie2016matching, alonso2006unifying}

\newpage

\begin{appendix}
\centerline{\sc Appendix: Proofs of Results}

\noindent{\bf Proof of Lemma \ref{lemma1}:} To prove this result we show that we cannot infer from the joint distribution of  $(W_i,X_i,G_i,Y^\pps_i,Y^\pp_i{\bf 1}_{G_i=\ppo})$, in combination with the assumptions, the distribution of $Y^\pp_i(1)$ conditional on $X_i$ and $G_i=\ppe$.
This distribution
 can be written as
\[ f_{Y^\pp(1)|X,G=\ppe}(y|x)=  f_{Y^\pp(1)|X,G=\ppe,W=1}(y|x) p(W=1|X=x,G=\ppe)\]
\[\hskip2cm +
 f_{Y^\pp(1)|X,G=\ppe,W=0}(y|x) p(W=0|X=x,G=\ppe).
\]
The data are not informative about the distribution of $Y^\pp_i(1)$ given $W_i=0,$ $X_i$ and $G_i=\ppe$. Assumption \ref{assumption:conditional} implies that this distribution is the same as the distribution of 
 $Y^\pp_i(1)$ given $W_i=0,$ $X_i$ and $G_i=\ppo$, but the data are not informative about that either. $\square$

\noindent{\bf Proof of Lemma \ref{lemma2}:}  To prove the result we show that
\[ G_i\ \perp\!\!\!\perp\ Y^\pps_i(1)\ \Bigl|\ X_i=x,W_i=1.\]
We can factor  the conditional distribution of $(Y^\pps_i(1),G_i)$ given $X_i$ and $W_i=1$  as
\[ f(Y^\pps(1),G|X,W=1)=f(Y^\pps(1)|G,X,W) f(G|X,W=1).\]
By the unconfoundedness assumptions, Assumptions \ref{assumption:random} and \ref{assumption:unconfoundedness} it follows that this is equal to
\[ f(Y^\pps(1)|G,X) f(G|X,W=1).\]
By Conditional External Validity (Assumption \ref{assumption:conditional}) this is equal to
\[ f(Y^\pps(1)|X) f(G|X,W=1).\]
By Assumptions  \ref{assumption:random} and \ref{assumption:unconfoundedness} this is equal to
\[ f(Y^\pps(1)|X,W=1) f(G|X,W=1),\]  which implies the conditional independence we set out to prove.
$\square$

\noindent{\bf Proof of Theorem \ref{theorem_main}:}\footnote{We are grateful to Nathan Kallus and Xiaojie Mao for pointing out a mistake in an earlier version of the proof of this theorem.}
To be clear here, we index the expectations operator by the random variable that the expectation is taken over. By definition
\[  \tau^\pp_\ppo=
\mme_{Y_i^\pp(1),Y_i^\pp(0)}\left[ \left.Y_i^\pp(1)-Y_i^\pp(0)\right| G_i=\ppo\right]=
\mme_{Y_i^\pp(1)}\left[ \left.Y_i^\pp(1)\right| G_i=\ppo\right]
-\mme_{Y_i^\pp(0)}\left[ \left.Y_i^\pp(0)\right| G_i=\ppo\right].\]
We focus on identification of the first term, which by iterated expectations can be written as
\begin{equation}
\label{een}\mme_{Y_i^\pp(1)}\left[ \left.Y_i^\pp(1)\right| G_i=\ppo\right]= \mme_{X_i}\left[\left.\mme_{Y_i^\pp(1)}\left[\left.Y^\pp_i(1)\right|X_i,G_i=\ppo\right]\right| G_i=\ppo\right].\end{equation}
Identification of the second term follows by the same argument.
By Conditional External Validity (Assumption \ref{assumption:conditional}), we can write the inner expectation as
\[\mme_{Y_i^\pp(1)}\left[\left.Y^\pp_i(1)\right|X_i,G_i=\ppo\right]=
\mme_{Y_i^\pp(1)}\left[\left.Y^\pp_i(1)\right|X_i,G_i=\ppe\right]
,\]
so that
(\ref{een}) is equal to
\begin{equation}
\label{twee} \mme_{X_i}\left[\left.\mme_{Y_i^\pp(1)}\left[\left.Y^\pp_i(1)\right|X_i,G_i=\ppe\right]\right| G_i=\ppo\right].\end{equation}
By iterated expectations this is equal to
\begin{equation}
\label{drie} \mme_{X_i}\left[\left.\mme_{Y_i^\pps(1)}\left[\left.\mme_{Y_i^\pp(1)}\left[\left.Y^\pp_i(1)\right| Y_i^\pps(1),X_i,G_i=\ppe\right]\right|X_i,G_i=\ppe\right]\right| G_i=\ppo\right].\end{equation}
By Conditional External Validity (Assumption \ref{assumption:conditional}), this is equal to
\begin{equation}
\label{vier} \mme_{X_i}\left[\left.\mme_{Y_i^\pps(1)}\left[\left.\mme_{Y_i^\pp(1)}\left[\left.Y^\pp_i(1)\right| Y_i^\pps(1),X_i,G_i=\ppo\right]\right|X_i,G_i=\ppe\right]\right| G_i=\ppo\right].\end{equation}
By Latent Unconfoundedness (Assumption \ref{assumption:new}) this is equal to
\begin{equation}
\label{vier} \mme_{X_i}\left[\left.\mme_{Y_i^\pps(1)}\left[\left.\mme_{Y_i^\pp(1)}\left[\left.Y^\pp_i(1)\right| Y_i^\pps(1),W_i=1,X_i,G_i=\ppo\right]\right|X_i,G_i=\ppe\right]\right| G_i=\ppo\right].\end{equation}
By the definitions $Y^\pp_i=Y^\pp_i(W_i)$ and $Y^\pps_i=Y^\pps_i(W_i)$ this is equal to
\begin{equation}
\label{vijf} \mme_{X_i}\left[\left.\mme_{Y_i^\pps(1)}\left[\left.\mme_{Y_i^\pp(1)}\left[\left.Y^\pp_i\right| Y_i^\pps,W_i=1,X_i,G_i=\ppo\right]\right|X_i,G_i=\ppe\right]\right| G_i=\ppo\right].\end{equation}
Define
\[ h(y^\pps,x)\equiv \mme_{Y_i^\pp(1)}\left[\left.Y^\pp_i\right| Y_i^\pps=y^\pps,W_i=1,X_i=x,G_i=\ppo\right],\]
so that (\ref{vijf}) is
\begin{equation}
\label{zes} \mme_{X_i}\left[\left.\mme_{Y_i^\pps(1)}\left[\left.
h(Y^\pps_i(1),X_i)
\right|X_i,G_i=\ppe\right]\right| G_i=\ppo\right].\end{equation}
Note that $h(y^\pps,x)$ is directly identified from the observational sample.

Because of the unconfoundedness in the experimental sample (Assumption \ref{assumption:random}), (\ref{zes}) is equal to
\begin{equation}
\label{zeven} \mme_{X_i}\left[\left.\mme_{Y_i^\pps(1)}\left[\left.
h(Y^\pps_i(1),X_i)
\right|W_i=1,X_i,G_i=\ppe\right]\right| G_i=\ppo\right].\end{equation}
By the definition of $Y^\pps_i=Y^\pps_i(W_i)$,
and because the conditional distribution of $Y^\pps_i(1)$ conditional on $W_i=1,X_i,G_i=\ppo$ is the same as the conditional distribution of  
 $Y^\pps_i$ conditional on $W_i=1,X_i,G_i=\ppo$, we can change the random variable that the expectation is taken over and write this as
\begin{equation}\label{acht} \mme_{X_i}\left[\left.\mme_{Y_i^\pps}\left[\left.
h(Y^\pps_i,X_i)
\right|W_i=1,X_i,G_i=\ppe\right]\right| G_i=\ppo\right].\end{equation}
The inner expectation
\[ k(x)\equiv \mme_{Y_i^\pps}\left[\left.
h(Y^\pps_i,X_i)
\right|W_i=1,X_i=x,G_i=\ppe\right],\]
is identified from the experimental sample.
The expectation 
\[ \mme[k(X_i)|G_i=\ppo],\]
is identified from the observational sample, which completes the proof.
$\square$

\end{appendix}

\newpage

\bibliographystyle{plainnat}
\bibliography{references}

\begin{thebibliography}{62}
\providecommand{\natexlab}[1]{#1}
\providecommand{\url}[1]{\texttt{#1}}
\expandafter\ifx\csname urlstyle\endcsname\relax
  \providecommand{\doi}[1]{doi: #1}\else
  \providecommand{\doi}{doi: \begingroup \urlstyle{rm}\Url}\fi

\bibitem[Adams et~al.(2006)Adams, Schatzkin, Harris, Kipnis, Mouw,
  Ballard-Barbash, Hollenbeck, and Leitzmann]{adams2006overweight}
Kenneth~F Adams, Arthur Schatzkin, Tamara~B Harris, Victor Kipnis, Traci Mouw,
  Rachel Ballard-Barbash, Albert Hollenbeck, and Michael~F Leitzmann.
\newblock Overweight, obesity, and mortality in a large prospective cohort of
  persons 50 to 71 years old.
\newblock \emph{New England Journal of Medicine}, 355\penalty0 (8):\penalty0
  763--778, 2006.

\bibitem[Alonso et~al.(2006)Alonso, Molenberghs, Geys, Buyse, and
  Vangeneugden]{alonso2006unifying}
Ariel Alonso, Geert Molenberghs, Helena Geys, Marc Buyse, and Tony
  Vangeneugden.
\newblock A unifying approach for surrogate marker validation based on
  prentice's criteria.
\newblock \emph{Statistics in Medicine}, 25\penalty0 (2):\penalty0 205--221,
  2006.

\bibitem[Angrist and Pischke(2008)]{angrist2008mostly}
Joshua~D Angrist and J{\"o}rn-Steffen Pischke.
\newblock \emph{Mostly harmless econometrics: An empiricist's companion}.
\newblock Princeton University Press, 2008.

\bibitem[Athey and Imbens(2006)]{athey2006identification}
Susan Athey and Guido~W Imbens.
\newblock Identification and inference in nonlinear difference-in-differences
  models.
\newblock \emph{Econometrica}, 74\penalty0 (2):\penalty0 431--497, 2006.

\bibitem[Athey and Wager(2017)]{athey2017efficient}
Susan Athey and Stefan Wager.
\newblock Efficient policy estimation.
\newblock \emph{arXiv preprint arXiv:1702.02896}, 2017.
\newblock URL \url{https://arxiv.org/abs/1702.02896}.

\bibitem[Athey et~al.(2019)Athey, Chetty, Imbens, and Kang]{athey2019surrogate}
Susan Athey, Raj Chetty, Guido~W Imbens, and Hyunseung Kang.
\newblock The surrogate index: Combining short-term proxies to estimate
  long-term treatment effects more rapidly and precisely.
\newblock Technical report, National Bureau of Economic Research, 2019.

\bibitem[Bhattacharya(2013)]{bhattacharya2013evaluating}
Debopam Bhattacharya.
\newblock Evaluating treatment protocols using data combination.
\newblock \emph{Journal of Econometrics}, 173\penalty0 (2):\penalty0 160--174,
  2013.

\bibitem[Biasi et~al.(2025)Biasi, Lafortune, and
  Sch{\"o}nholzer]{biasi2025works}
Barbara Biasi, Julien Lafortune, and David Sch{\"o}nholzer.
\newblock What works and for whom? effectiveness and efficiency of school
  capital investments across the u.s.
\newblock \emph{The Quarterly Journal of Economics}, page qjaf013, 2025.

\bibitem[Bleemer(2022)]{bleemer2022affirmative}
Zachary Bleemer.
\newblock Affirmative action, mismatch, and economic mobility after
  california’s proposition 209.
\newblock \emph{The Quarterly Journal of Economics}, 137\penalty0 (1):\penalty0
  115--160, 2022.

\bibitem[Card(1990)]{cardmariel}
David Card.
\newblock The impact of the mariel boatlift on the miami labor market.
\newblock \emph{Industrial and Labor Relations Review}, 43\penalty0
  (2):\penalty0 245--257, 1990.

\bibitem[Card and Krueger(1994)]{cardkrueger1}
David Card and Alan Krueger.
\newblock Minimum wages and employment: Case study of the fast-food industry in
  new jersey and pennsylvania.
\newblock \emph{American Economic Review}, 84\penalty0 (4):\penalty0 772--793,
  1994.

\bibitem[Cascio and Staiger(2012)]{cascio2012knowledge}
Elizabeth~U Cascio and Douglas~O Staiger.
\newblock Knowledge, tests, and fadeout in educational interventions.
\newblock Technical report, National Bureau of Economic Research, 2012.

\bibitem[Chen and Ritzwoller(2023)]{chen2023semiparametric}
Jiafeng Chen and David~M Ritzwoller.
\newblock Semiparametric estimation of long-term treatment effects.
\newblock \emph{Journal of Econometrics}, 237\penalty0 (2):\penalty0 105545,
  2023.

\bibitem[Chen and Santos(2018)]{chen2018overidentification}
Xiaohong Chen and Andres Santos.
\newblock Overidentification in regular models.
\newblock \emph{Econometrica}, 86\penalty0 (5):\penalty0 1771--1817, 2018.

\bibitem[Chernozhukov et~al.(2022)Chernozhukov, Newey, and
  Singh]{chernozhukov2022automatic}
Victor Chernozhukov, Whitney~K Newey, and Rahul Singh.
\newblock Automatic debiased machine learning of causal and structural effects.
\newblock \emph{Econometrica}, 90\penalty0 (3):\penalty0 967--1027, 2022.

\bibitem[Chetty et~al.(2011)Chetty, Friedman, Hilger, Saez, Schanzenbach, and
  Yagan]{chetty2011does}
Raj Chetty, John~N Friedman, Nathaniel Hilger, Emmanuel Saez, Diane~Whitmore
  Schanzenbach, and Danny Yagan.
\newblock How does your kindergarten classroom affect your earnings? evidence
  from project star.
\newblock \emph{The Quarterly Journal of Economics}, 126\penalty0 (4):\penalty0
  1593--1660, 2011.

\bibitem[Chetty et~al.(2014)Chetty, Friedman, and
  Rockoff]{chetty2014onemeasuring}
Raj Chetty, John~N Friedman, and Jonah~E Rockoff.
\newblock Measuring the impacts of teachers i: Evaluating bias in teacher
  value-added estimates.
\newblock \emph{American Economic Review}, 104\penalty0 (9):\penalty0
  2593--2632, 2014.

\bibitem[D'Agostino et~al.(2006)D'Agostino, Campbell, and
  Greenhouse]{d2006surrogate}
Ralph~B D'Agostino, Michael~J Campbell, and Joel~B Greenhouse.
\newblock Surrogate markers: back to the future.
\newblock \emph{Statistics in medicine}, 25\penalty0 (2):\penalty0 181--182,
  2006.

\bibitem[Dehejia(2005)]{dehejia2005program}
Rajeev~H Dehejia.
\newblock Program evaluation as a decision problem.
\newblock \emph{Journal of Econometrics}, 125\penalty0 (1):\penalty0 141--173,
  2005.

\bibitem[Deming(2009)]{deming2009early}
David Deming.
\newblock Early childhood intervention and life-cycle skill development:
  Evidence from head start.
\newblock \emph{American Economic Journal: Applied Economics}, 1\penalty0
  (3):\penalty0 111--134, 2009.

\bibitem[Gupta et~al.(2019)Gupta, Kohavi, Tang, Xu, Andersen, Bakshy, Cardin,
  Chandran, Chen, Coey, et~al.]{gupta2019top}
Somit Gupta, Ronny Kohavi, Diane Tang, Ya~Xu, Reid Andersen, Eytan Bakshy,
  Niall Cardin, Sumita Chandran, Nanyu Chen, Dominic Coey, et~al.
\newblock Top challenges from the first practical online controlled experiments
  summit.
\newblock \emph{ACM SIGKDD Explorations Newsletter}, 21\penalty0 (1):\penalty0
  20--35, 2019.

\bibitem[Heckman and Robb(1985)]{heckman1985alternative}
James Heckman and Richard Robb.
\newblock Alternative methods for evaluating the impact of interventions: An
  overview.
\newblock \emph{Journal of Econometrics}, 30\penalty0 (1-2):\penalty0 239--267,
  1985.

\bibitem[Heckman(1979)]{heckman1979sample}
James~J Heckman.
\newblock Sample selection bias as a specification error.
\newblock \emph{Econometrica}, 47\penalty0 (1):\penalty0 153--161, 1979.

\bibitem[Heckman et~al.(2006)Heckman, Stixrud, and Urzua]{heckman2006effects}
James~J Heckman, Jora Stixrud, and Sergio Urzua.
\newblock The effects of cognitive and noncognitive abilities on labor market
  outcomes and social behavior.
\newblock \emph{Journal of Labor Economics}, 24\penalty0 (3):\penalty0
  411--482, 2006.

\bibitem[Hirano and Porter(2009)]{hirano2009asymptotics}
Keisuke Hirano and Jack~R Porter.
\newblock Asymptotics for statistical treatment rules.
\newblock \emph{Econometrica}, 77\penalty0 (5):\penalty0 1683--1701, 2009.

\bibitem[Hotz et~al.(2005)Hotz, Imbens, and Mortimer]{hotz2005predicting}
V~Joseph Hotz, Guido~W Imbens, and Julie~H Mortimer.
\newblock Predicting the efficacy of future training programs using past
  experiences at other locations.
\newblock \emph{Journal of Econometrics}, 125\penalty0 (1):\penalty0 241--270,
  2005.

\bibitem[Imbens(2000)]{imbens2000}
Guido Imbens.
\newblock The role of the propensity score in estimating dose--response
  functions.
\newblock \emph{Biometrika}, 87\penalty0 (0):\penalty0 706--710, 2000.

\bibitem[Imbens et~al.(2025)Imbens, Kallus, Mao, and Wang]{imbens2025long}
Guido Imbens, Nathan Kallus, Xiaojie Mao, and Yuhao Wang.
\newblock Long-term causal inference under persistent confounding via data
  combination.
\newblock \emph{Journal of the Royal Statistical Society Series B: Statistical
  Methodology}, 87\penalty0 (2):\penalty0 362--388, 2025.

\bibitem[Imbens and Newey(2009)]{imbens2009identification}
Guido~W Imbens and Whitney~K Newey.
\newblock Identification and estimation of triangular simultaneous equations
  models without additivity.
\newblock \emph{Econometrica}, 77\penalty0 (5):\penalty0 1481--1512, 2009.

\bibitem[Imbens and Rubin(2015)]{imbens2015causal}
Guido~W Imbens and Donald~B Rubin.
\newblock \emph{Causal Inference in Statistics, Social, and Biomedical
  Sciences}.
\newblock Cambridge University Press, 2015.

\bibitem[Imbens and Wooldridge(2009)]{imbens2009recent}
Guido~W Imbens and Jeffrey~M Wooldridge.
\newblock Recent developments in the econometrics of program evaluation.
\newblock \emph{Journal of Economic Literature}, 47\penalty0 (1):\penalty0
  5--86, 2009.

\bibitem[Kallus and Mao(2020)]{kallus2020role}
Nathan Kallus and Xiaojie Mao.
\newblock On the role of surrogates in the efficient estimation of treatment
  effects with limited outcome data.
\newblock \emph{arXiv preprint arXiv:2003.12408}, 2020.

\bibitem[Kallus et~al.(2018)Kallus, Puli, and Shalit]{kallus2018removing}
Nathan Kallus, Aahlad~Manas Puli, and Uri Shalit.
\newblock Removing hidden confounding by experimental grounding.
\newblock \emph{Advances in Neural Information Processing Systems}, 31, 2018.

\bibitem[Kane and Staiger(2008)]{kane2008estimating}
Thomas~J Kane and Douglas~O Staiger.
\newblock Estimating teacher impacts on student achievement: An experimental
  evaluation.
\newblock Technical report, National Bureau of Economic Research, 2008.

\bibitem[Kline and Walters(2019)]{kline2019heckits}
Patrick Kline and Christopher~R Walters.
\newblock On heckits, late, and numerical equivalence.
\newblock \emph{Econometrica}, 87\penalty0 (2):\penalty0 677--696, 2019.

\bibitem[Krueger(1999)]{krueger1999experimental}
Alan~B Krueger.
\newblock Experimental estimates of education production functions.
\newblock \emph{The Quarterly Journal of Economics}, 114\penalty0 (2):\penalty0
  497--532, 1999.

\bibitem[Little and Rubin(2019)]{little2019statistical}
Roderick~JA Little and Donald~B Rubin.
\newblock \emph{Statistical analysis with missing data}, volume 793.
\newblock Wiley, 2019.

\bibitem[Manski(1990)]{manski_bounds}
Charles~F Manski.
\newblock Nonparametric bounds on treatment effects.
\newblock \emph{The American Economic Review}, 80\penalty0 (2):\penalty0
  319--323, 1990.

\bibitem[Manski(2004)]{manski2004statistical}
Charles~F Manski.
\newblock Statistical treatment rules for heterogeneous populations.
\newblock \emph{Econometrica}, 72\penalty0 (4):\penalty0 1221--1246, 2004.

\bibitem[Mariano et~al.(2024)Mariano, Martorell, and
  Berglund]{mariano2024effects}
Louis~T Mariano, Paco Martorell, and Tiffany Berglund.
\newblock The effects of grade retention on high school outcomes: Evidence from
  new york city schools.
\newblock \emph{Journal of Research on Educational Effectiveness}, pages 1--31,
  2024.

\bibitem[Mealli and Pacini(2013)]{mealli2013using}
Fabrizia Mealli and Barbara Pacini.
\newblock Using secondary outcomes and covariates to sharpen inference in
  instrumental variable settings.
\newblock \emph{Journal of the American Statistical Association}, 108:\penalty0
  1120--1131, 2013.

\bibitem[Meza and Singh(2024)]{meza2024}
Isaac Meza and Rahul Singh.
\newblock Nested nonparametric instrumental variable regression: Long term,
  mediated, and time varying treatment effects, 2024.
\newblock URL \url{https://arxiv.org/abs/2112.14249}.

\bibitem[Mogstad and Torgovitsky(2018)]{mogstad2018identification}
Magne Mogstad and Alexander Torgovitsky.
\newblock Identification and extrapolation of causal effects with instrumental
  variables.
\newblock \emph{Annual Review of Economics}, 10:\penalty0 577--613, 2018.

\bibitem[Mogstad et~al.(2018)Mogstad, Santos, and
  Torgovitsky]{mogstad2018using}
Magne Mogstad, Andres Santos, and Alexander Torgovitsky.
\newblock Using instrumental variables for inference about policy relevant
  treatment parameters.
\newblock \emph{Econometrica}, 86\penalty0 (5):\penalty0 1589--1619, 2018.

\bibitem[{New York State Education Department}(2025)]{nysed_downloads}
{New York State Education Department}.
\newblock {NYSED Data Site: Downloads}.
\newblock \url{https://data.nysed.gov/downloads.php}, 2025.
\newblock Accessed: 2025-04-25.

\bibitem[Newey(1994)]{newey1994asymptotic}
Whitney~K Newey.
\newblock The asymptotic variance of semiparametric estimators.
\newblock \emph{Econometrica}, pages 1349--1382, 1994.

\bibitem[Park and Sasaki(2024{\natexlab{a}})]{park2024bracketing}
Yechan Park and Yuya Sasaki.
\newblock A bracketing relationship for long-term policy evaluation with
  combined experimental and observational data.
\newblock \emph{arXiv preprint arXiv:2401.12050}, 2024{\natexlab{a}}.

\bibitem[Park and Sasaki(2024{\natexlab{b}})]{park2024informativeness}
Yechan Park and Yuya Sasaki.
\newblock The informativeness of combined experimental and observational data
  under dynamic selection.
\newblock \emph{arXiv preprint arXiv:2403.16177}, 2024{\natexlab{b}}.

\bibitem[Pearl et~al.(2014)Pearl, Bareinboim, et~al.]{pearl2014external}
Judea Pearl, Elias Bareinboim, et~al.
\newblock External validity: From do-calculus to transportability across
  populations.
\newblock \emph{Statistical Science}, 29\penalty0 (4):\penalty0 579--595, 2014.

\bibitem[Prentice(1989)]{prentice1989surrogate}
Ross~L Prentice.
\newblock Surrogate endpoints in clinical trials: definition and operational
  criteria.
\newblock \emph{Statistics in Medicine}, 8\penalty0 (4):\penalty0 431--440,
  1989.

\bibitem[Ridder and Moffitt(2007)]{ridder2007econometrics}
Geert Ridder and Robert Moffitt.
\newblock The econometrics of data combination.
\newblock \emph{Handbook of Econometrics}, 6:\penalty0 5469--5547, 2007.

\bibitem[Rosenbaum and Rubin(1983)]{rosenbaum1983central}
Paul~R Rosenbaum and Donald~B Rubin.
\newblock The central role of the propensity score in observational studies for
  causal effects.
\newblock \emph{Biometrika}, 70\penalty0 (1):\penalty0 41--55, 1983.

\bibitem[Rosenman et~al.(2018)Rosenman, Owen, Baiocchi, and
  Banack]{rosenman2018propensity}
Evan Rosenman, Art~B Owen, Michael Baiocchi, and Hailey Banack.
\newblock Propensity score methods for merging observational and experimental
  datasets.
\newblock \emph{arXiv preprint arXiv:1804.07863}, 2018.

\bibitem[Rosenman et~al.(2020)Rosenman, Basse, Owen, and
  Baiocchi]{rosenman2020combining}
Evan Rosenman, Guillaume Basse, Art Owen, and Michael Baiocchi.
\newblock Combining observational and experimental datasets using shrinkage
  estimators.
\newblock \emph{arXiv preprint arXiv:2002.06708}, 2020.

\bibitem[Rubin(1974)]{rubin1974estimating}
Donald~B Rubin.
\newblock Estimating causal effects of treatments in randomized and
  nonrandomized studies.
\newblock \emph{Journal of Educational Psychology}, 66\penalty0 (5):\penalty0
  688, 1974.

\bibitem[Rubin(1976)]{rubin1976inference}
Donald~B Rubin.
\newblock Inference and missing data.
\newblock \emph{Biometrika}, 63\penalty0 (3):\penalty0 581--592, 1976.

\bibitem[Rubin(1987)]{rubin2004multiple}
Donald~B. Rubin.
\newblock \emph{Multiple Imputation for Nonresponse in Surveys}.
\newblock Wiley Series in Probability and Mathematical Statistics. Wiley, New
  York, 1987.
\newblock ISBN 047108705X, 9780471087052.

\bibitem[Shadish et~al.(2002)Shadish, Cook, and Campbell]{shadishcookcampbell}
William~R Shadish, Thomas~D Cook, and Donald~T Campbell.
\newblock \emph{Experimental and quasi-experimental designs for generalized
  causal inference.}
\newblock Houghton, Mifflin and Company, 2002.

\bibitem[Wooldridge(2015)]{wooldridge2015control}
Jeffrey~M Wooldridge.
\newblock Control function methods in applied econometrics.
\newblock \emph{Journal of Human Resources}, 50\penalty0 (2):\penalty0
  420--445, 2015.

\bibitem[Wooldridge(2010)]{wooldridge2010econometric}
J.M. Wooldridge.
\newblock \emph{Econometric Analysis of Cross Section and Panel Data}.
\newblock The MIT Press. MIT Press, 2010.
\newblock ISBN 9780262232586.

\bibitem[Word et~al.(1990)Word, Johnston, Bain, Fulton, Zaharias, Achilles,
  Lintz, Folger, and Breda]{word1990state}
Elizabeth Word, John Johnston, Helen~P Bain, B~DeWayne Fulton, Jayne~B
  Zaharias, Charles~M Achilles, Martha~N Lintz, John Folger, and Carolyn Breda.
\newblock The state of tennessee’s student/teacher achievement ratio (star)
  project.
\newblock \emph{Tennessee Board of Education}, 1990.

\bibitem[Zhou et~al.(2018)Zhou, Athey, and Wager]{zhou2018offline}
Zhengyuan Zhou, Susan Athey, and Stefan Wager.
\newblock Offline multi-action policy learning: Generalization and
  optimization.
\newblock \emph{arXiv preprint arXiv:1810.04778}, 2018.

\end{thebibliography}

\end{document}